# IDGraphs: Intrusion Detection and Analysis Using Stream Compositing

Pin Ren
Yan Gao
Zhichun Li
Yan Chen
Benjamin Watson

Northwestern University
{p-ren,yga751,lizc,ychen,watsonb}@cs.northwestern.edu

**ABSTRACT**

Traffic anomalies and attacks are commonplace in today's networks and identifying them rapidly and accurately is critical for large network operators. For a statistical intrusion detection system (IDS), it is crucial to detect at the flow-level for accurate detection and mitigation. However, existing IDS systems offer only limited support for 1) interactively examining detected intrusions and anomalies, 2) analyzing worm propagation patterns, 3) and discovering correlated attacks. These problems are becoming even more acute as the traffic on today's high-speed routers continues to grow.

*IDGraphs* is an interactive visualization system for intrusion detection that addresses these challenges. The central visualization in the system is a flow-level trace plotted with time on the horizontal axis and aggregated number of unsuccessful connections on the vertical axis. We then summarize a stack of tens or hundreds of thousands of these traces using the *Histographs* [RW05] technique, which maps data frequency at each pixel to brightness. Users may then interactively query the summary view, performing analysis by highlighting subsets of the traces. For example, brushing a linked correlation matrix view highlights traces with similar patterns, revealing distributed attacks that are difficult to detect using standard statistical analysis.

We apply IDGraphs system to a real network router data-set with 179M flow-level records representing a total traffic of 1.16TB. The system successfully detects and analyzes a variety of attacks and anomalies, including port scanning, worm outbreaks, stealthy TCP SYN floodings, and some distributed attacks.

**CR Categories:** C.2.0 [Computer-Communication Networks]: General—security and protection; H.5.2 [Information Systems]: Information Interfaces and Presentation—User Interfaces; K.6.5 [Management of Computing and Information Systems]: Security and Protection—invasive software

## 1 Introduction

Traffic anomalies and attacks are commonplace in today's networks. It is estimated that malicious code (viruses, worms and Trojan horses) caused over $28 billion in economic losses in

2003, and will grow to over $75 billion by 2007[1]. For these reasons, large network operators place great importance on rapid and accurate identification of traffic anomalies and attacks.

Most existing intrusion detection systems (IDSs) identify attacks using specific patterns in the attack traffic called signatures. But such IDSs cannot detect unknown network attacks, and attackers can easily foil detection by garbling their signatures. Other statistical IDSs [MVS01] [WZS04] use overall traffic to detect attacks, but suffer from inaccuracies and difficulties in finding attack flows, even when anomalies are correctly identified. There are also a few flow-level detection schemes [PAX99] [Roe01], which keep status for specific flows, but the following questions remain open.

- Do intrusions such as TCP SYN flooding and port scans have characteristic time series patterns, when observed from edge network routers? For instance, are there any common patterns for spread of a specific worm that might indicate its propagation strategy? Answering these questions will be difficult to obtain without visualization, especially with today's huge network flows.

- How can we identify correlated attacks, especially when they are new? This is a difficult challenge for the intrusion detection (ID) community. To the best of our knowledge, almost all systems have to treat attacks independently, even after detecting the attacks.

- How can discovered intrusions and anomalies be analyzed interactively? One of the key challenges for statistical detection is the threshold for attacks. How will the attacks and their distributions/patterns change when we change the detection threshold?

*IDGraphs* is an interactive visualization system designed to address these challenges, supporting intrusion detection over massive network traffic streams. It has the following features:

- A novel data-to-space mapping for discovery of attack patterns. We plot the number of unsuccessful connections (SYN-SYN/ACK) vs. time in our graphs. We are suspicious of any connections that fail too frequently. For detection of TCP SYN flooding, we use time series corresponding to unique destination IP (DIP) and port (Dport) pairs. For detection of horizontal scans, series correspond to source IP (SIP,Dport) keys, and for vertical scan detection, to (SIP,DIP) keys. Other series keys are also possible.

- High visual scalability through the use of Histographs [RW05]. Tens or hundreds of thousands of time series can be viewed at once, with frequency of network events indicated by pixel brightness.

- A linked correlation matrix view that reveals correlated attacks. Brushing reveals correlated time series patterns. To the best of our knowledge, we are the first to use such views for intrusion detection.

- A search and filter interface for ungraphed network data dimensions such as SIP and Dport.

We demonstrate IDGraphs on a single day of NetFlow network traffic traces collected at edge routers at Northwestern University, which has several OC-3 links. These traces totalled 179M records and 1.16TB of traffic. IDGraphs reveals the port scanning of virus and worm propagation, the pattern of stealthy TCP SYN flooding, as well as the correlated action of distributed attacks.

---

[1] *Email Defense Statistics* http://www.mxlogic.com/PDFs/IndustryStats.2.28.04.pdf

# 2 Previous Work

## 2.1 Intrusion Detection Systems

An IDS is a type of security management system for computers and networks. It gathers and analyzes information from various areas within a computer or a network to identify possible security breaches, which include both intrusions and misuse. With the rapid growth of network bandwidth and fast emergence of new attacks/worms, network IDSs have drawn more attention from researchers.

Many network IDSs like Bro [PAX99] and Snort [Roe01] check packet payload for virus/worm signatures. However, such schemes do not scale to high-speed network links. To detect large scale attacks, many researchers have proposed techniques based on the statistical characteristics of the intrusions.

We classify these techniques into two rough categories: 1)detection based on overall traffic [MVS01] [WZS04] such as Change Point Monitor (CPM), which tends to be inaccurate and cannot find real attack flows; and 2) flow-level detection [PAX99] [Roe01] such as Threshold Random Walk (TRW), which is vulnerable to denial-of-service (DoS) attacks with randomly spoofed IP addresses. Flow-level detection is especially vulnerable on high-speed networks, since the sequential hypothesis testing scheme it uses needs to maintain a per-SIP table for detection. Gao et al. [GLC05] recently addressed this problem using a reversible sketch technique.

Most ID technologies perform detection on individual traffic flows, rather than looking for the correlations between multiple flows. These methods can only provide a small snapshot of globally distributed attacks. More recently developed correlation information analyses [ATS03] [KTK01] address this problem, reducing the high volume of alerts and false positives.

## 2.2 Visualization For Internet Security

In applying interactive visualization to Internet security research, researchers exploit the innate and highly efficient human ability to process visual information, enabling the complex tasks of network security monitoring and intrusion detection to be performed in an accurate and timely manner. Many systems [TMW02][LAU04][YYT04][MMK04][LYL04] have addressed this problem. All of them provide interactive visual support for anomaly detection.

PortVis [MMK04] produces visualizations of network traffic using 2D plots with time and port number as axes, and summarizing the network activity at each location in the plot (a time/port pair) using color. Users can drill down to display traffic information at finer temporal and port resolutions.

VisFlowConnect [YYT04] uses a simple application of parallel coordinates \cite{PC90} to display incoming and outgoing network flow data as links between two machines or domains. (Parallel coordinates are a widely used technique for plotting high-dimensional data). It also employs a variety of visual cues to help detect attacks.

The Spinning Cube [LAU04] maps SIP, DIP and Dport to the axes in a 3D plot. The amount of network activity is visualized interactively in the plot using color, displaying certain attacks (especially port scans) very clearly.

NVisionIP [LYL04] visualizes network flow data in a 2D matrix with IP addresses on each axis. Each cell in the matrix represents the interaction between the corresponding network hosts. Users can reduce or increase detail in the current view.

## 3 The Threat Model And Data Collection

### 3.1 Threat Model

Ultimately, we want to detect as many attacks as possible. As a first step, we focus on arguably the two most popular intrusions for detection: DoS TCP flooding attacks\footnote{According to the CERT DoS threat model[2], DoS attacks may also include corruption attacks, which are excluded here because they are often application/protocol specific.} and port scans (mostly for worm propagation). It is reported that more than 90% of DoS attacks are TCP SYN flooding attacks [WZS04].

*Figure 1 about here*

Scans are probably the most common and versatile type of intrusion. Based on source/dest IP and the port number involved in the scans, there are three well known types of scans: horizontal scan, vertical scan, and block scan [SHM02]. The classification is illustrated in Figure 1.Unlike DoS attacks, the attacker needs to use a real source IP address, since he/she needs to see the result of the scan in order to know what ports are actually open [SHM02]. Horizontal scans are the most common type of scan, and scans certain ports across an interesting range of IP addresses. The port number is often unique because it reflects the vulnerability the virus/worm or attackers try to exploit. A vertical scan is a scan of some or all ports on a single host, with the rationale that the attacker is interested in this particular host, and wishes to characterize its active services to find which exploits to attempt [SHM02]. The third type of scan, a block scan, is a combination of horizontal and vertical scans over numerous services on numerous hosts [SHM02].

| *the destination IP* | **DIP** |
|---|---|
| *the source IP* | **SIP** |
| *the destination port* | **Dport** |
| *the source port* | **Sport** |

*Table 1: The fields in IP header that we may use in detection*

### 3.2 Data Collection

Our system is based on preprocessed NetFlow data, but it is easy to extend to other data sources. NetFlow data was originally derived from Cisco routers caching recent flows for lookup efficiency, and it has now become the de facto standard for router traffic monitoring, accepted by all other major router vendors. NetFlow is identified as a unidirectional stream of packets between a given source and destination, both of which are defined by a network-layer IP address and transport-layer source and destination port numbers. Here we only consider the attacks in TCP protocol, in other words, the TCP SYN Flooding attacks and TCP port scans. We analyze the attributes in TCP/IP headers and select a small set of metrics for flow-level traffic monitoring, the possible fields we can use are shown in Table 1. Normally, attackers can choose

---

[2] According to the CERT DoS threat model: http://www.cert.org/tech_tips/denial_of_service.html, DoS attacks may also include corruption attacks, which are excluded here because they are often application/protocol specific.

TCP source ports arbitrarily, so Sport may not be a good metric for attack detection. For the other three fields, we could consider all the combinations of these three fields, but the key ( SIP, DIP,Dport) can only find non-spoofed SYN flooding, so we do not use it in detection. Table 2 shows the other combinations and their selectivity to different types of attacks. Here, we define the selectivity of a key as the capability of differentiating between different types of attacks.

| Types of Keys | SYN flooding | hscan | vscan | bscan |
| --- | --- | --- | --- | --- |
| (SIP,Dport) | Part(non-spoofed) | Yes | No | Yes |
| (DIP,Dport) | Yes | No | No | No |
| (SIP,DIP) | Part(non-spoofed) | No | Yes | Yes |
| (SIP) | Part(non-spoofed) | Yes | Yes | Yes |
| (DIP) | Yes | No | Yes | Yes |
| (Dport) | Yes | Yes | No | Yes |

*Table 2: The selectivity of different types of keys. The bottom three single-field keys are less selective. (hscan=horizontal scan; vscan=vertical scan; bscan=block scan.)*

## 4 The Design of IDGraphs

IDGraphs is built on top of the Histographs visualization system [RW05], with the enhancements designed specifically for visualizing NetFlow datasets. The data input can be any one of the three aggregated NetFlow data files we discussed above (Figure 2, 11 and 12). In preprocessing we sequence records by key and then time to form a time series for each key. We filter out streams with less than 5 unsuccessful connections over the whole time range.

*Figure 2 about here*

*IDGraphs* is designed to help Internet security experts inspect their NetFlow data visually and perform deep analysis. Users can quickly identify possible anomalies or attacks using overviews (Figure 2), then follow up with in-depth analyses by querying those possible anomalies (Figure 3). One such analysis is identifying consistent temporal patterns in anomalies, which users can perform in two different ways. Dynamic querying selects and highlights streams with the same or similar SIPs or DIPs (Figure 5). Linked correlation views (Figure 9, 10) help the user select highly correlated streams and highlight them in the main view.

*Figure 3 about here*

IDGraphs can also be used for interactive visual tuning of automated intrusion detection techniques. Detection thresholds can be investigated using a vertical slider that highlights all streams with a minimum number of unsuccessful connections. Users can then annotate (Figure 7) interesting data subsets for further analysis and presentation by the users themselves, or their collaborators.

### 4.1 Visual Mapping

In visualizing data, we must define a mapping from the data space to the screen space. Lau [LAU04] maps SIP, DIP, and Dport to the three axes of a cube. PortVis [MMK04] treats the Dport as a 2 byte number, and maps each byte to the axes of a 2D plot. VisFlowConnect [YYT04] VisFlowConnect shows incoming and outgoing links by mapping source and destination to parallel axes, and connecting them with edges.

Unlike previous systems, IDGraphs displays time series data, a temporally ordered sequence of SYN-SYN/ACK values for each file key. We therefore map time to the horizontal axis, and SYN-SYN/ACK to the vertical axis of a 2D plot. Users can also transform the data before this mapping, producing for example a log(SYN-SYN/ACK) mapping to the vertical axis that compresses the data and makes more efficient use of display space. We map log(0) to -1 (Figure 2).

This time series mapping quickly reveals temporal patterns in network flow. It also effectively maps importance to the vertical axis, since higher SYN-SYN/ACK values are more suspicious, and more likely to be intrusions. This highlights potential attacks for users. The number of streams can be viewed at once is only limited by available machine memory.

Since we visualize thousands of streams at once using this mapping, we face an occlusion problem: multiple data points can be mapped to the same display pixel. The base Histographs system [RW05] is designed for plotting dense and high-dimensional data by "stacking" or compositing graphs, addresses this problem with a number of techniques. First, similar to the Information Mural system [JS98], the number of data points at a pixel (frequency) is mapped to pixel luminance, darkening those regions of the plot where data is dense. This highlights the main data trends, but unfortunately, it also makes it difficult to perceive outliers. Histographs addresses this problem in two ways. First, it introduces a new, contrast-weighted mapping between data and luminance that highlights changes in data frequency. Second, when data points are isolated, it adds lower spatial frequencies to them to increase their visibility ("splatting") without adjusting the data-luminance mapping. These measures are particularly important in IDGraphs, where outliers are precisely what users are seeking.

These visual mappings provide an effective overview of the NetFlow data, while also revealing concurrent anomalous activity. Events such as virus outbreaks or port scanning will quickly attract attention from the user (Figure 2).

### 4.2 Interactive Query

Interaction is the key to performing deep analysis with IDGraphs. Our design is guided by Shneiderman's visual information-seeking Mantra [Shn96], aiming to provide detailed information whenever the user asks for it. The dynamic query techniques pioneered by Shneiderman also heavily influenced our design.

The ability to click and query is central to interactive analysis with IDGraphs. Users can click on any pixel to reveal a pop-up menu (Figure 3, 5) showing textual information about the data from different streams aggregated by this pixel. For the (SIP,Dport) file, this reveals SIP, Dport, and the SYN-SYN/ACK difference. Currently selected streams in the query interface are shown by color bars, which have the same color as the lines highlighting the streams in the IDGraph itself.

Clicking for selection is tolerant of inaccuracy, allowing a one-pixel mismatch between the location of nearby data and cursor location. This is especially effective when the user wants to query an isolated data pixel.

By selecting streams in the pop-up menu, users can highlight only those streams with certain keys. A shortcut button quickly selects and highlights all the streams in the list (Figure 5).

*Figure 4 about here*

We highlight streams in the IDGraph by linking the data points of each selected time series with lines. Different colors are applied to each stream. Stream data may not be contiguous; in such situations the streams appear as several disconnected polylines, with filled circles emphasizing the start and the end point of each trace (Figures 4, 6).

*Figure 5 about here*

Having found suspicious network activity, users will often try to generalize the discovery by searching for other streams with the similar features. To address this problem, we provide a more general query interface that allows users to select streams with the same or similar source and destination (Figure 5). Users need not click on a displayed IDGraph feature to use this interface. In Figure 6, we selected all the streams with destination port 3306.

*Figure 6 about here*

To provide real-time intrusion detection, IDS systems often use default detection thresholds to identify suspicious network activity. These thresholds are very important in both simulation and actual detection. Determining such thresholds is difficult. IDGraphs allows users to examine the effectiveness and impact of different thresholds with vertical slider brushing (Figure 4), which highlights supra-threshold streams interactively. Users can adjust the possible detection threshold interactively by moving the slider. As they do so, all the streams with at least one data value over the threshold will be selected and highlighted. This enables users to study the effect on detection of different thresholds conveniently and interactively, with immediate visual feedback.

*Figure 7 about here*

### 4.3 Data Space Zoom

Having seen an overview, users can zoom in on interesting data features, obtaining more detailed and finer scale information about the data set. To zoom, users draw a rectangle over a region of interest in an IDGraph, simultaneously selecting a time and failure count (SYN-SYN/ACK) range (Figure 8). For more precision, sliders may also be used. When the selection is complete, a new view appears, displaying the selected time/failure range, and any streams that intersect this range. Zooms may also be performed on zoomed views. To maintain a visual correspondence between a zoomed view and its parent context, we use the same frequency-to-luminance mapping from the parent view. To generate zoomed views, we organize netflow data into a detail pyramid, with higher temporal resolutions used at the finer, detailed levels.. We use the finest temporal resolution supported by display resolution.

*Figure 8 about here*

### 4.4 Correlation Analysis

To help users form and test hypotheses about relationships between two or more NetFlow streams, or simply to identify streams with similar temporal NetFlow signatures, IDGraphs provides a linked correlation matrix view (Figure 9 and 10). In this matrix, each row and column represents one stream, and displays correlation values to all other streams. In each matrix cell, red indicates negative correlation and green positive, while luminance increases with the absolute magnitude of the correlation. When the number of matrix cells is greater than the number of pixels within the designated display area, we perform necessary screen-space scaling to display the matrix and therefore each pixel may correspond to multiple streams. Using two sliders, users can interactively specify a time range over which to construct the matrix and examine correlations (Figure 10).

*Figures 9 and 10 about here*

Using brushing and linking, a standard information visualization technique, users can link the correlation view to the main IDGraph, with streams highlighted in one highlighted in the other. This provides two ways of visualizing and interacting with the same data. By "brushing" or selecting interesting data in one view, users can study the shape of the data in the linked view. In IDGraphs' linked correlation view, users can brush using a horizontal stroke or instead use two precise sliders. The corresponding set of NetFlow streams is highlighted in the main view for further attention (Figure 9 and 10).

This brushing and linking is not particularly useful when the correlated streams are distributed widely across the correlation view. We increase effectiveness by reordering NetFlow streams in the matrix into correlated clusters. To perform this reordering, we apply the correlation matrix ordering technique described by Friendly [Fri02]. Each row (column) in the matrix is treated as a point in a high-dimensional space, and principal component analysis is applied. Each row (column) is then projected into the 2D space described by the first two eigenvectors of the correlation matrix. These projected 2D points are then ordered radially, and same ordering applied to the rows (columns) of the correlation matrix.

## 5 Case Studies

In this section we describe several examples of the use of IDGraphs for anomaly detection.

### 5.1 A Horizontal Scan Caused by a Coordinated Worm Attack

Figure 2 is a visualization of five hours of NetFlow data organized into time series with (SIP,Dport) keys. In the middle of hour 3 there is a very suspicious vertical linear structure. By selecting the streams that reach its SYN-SYN/ACK range (Figure 4), we reveal many short streams with almost no unsuccessful connections outside of the time range spanned by this linear structure, yet with a sudden increase in unsuccessful connections at the time of this structure. Clicking on these streams reveals that they are from different hosts (SIPs), but communicate with 3 common destination ports: 5554, 9898 and 1023. These are ports targeted by the Dabber backdoor and Sasse worm. We discovered these coordinated attacks without prior knowledge of this port information. Having identified these suspicious ports, we can select the streams connecting to those ports via the query interface shown in Figure 5, quickly identifying all the possible attacks by this worm within our dataset, even if they are smaller and stealthier. Because they are highly similar, these streams are also salient in the correlation matrix view (Figure 9), appearing as the large green block in the middle of the matrix.

### 5.2 A Block Scan and Temporal Similarities in Horizontal Scans

Those streams with a high number of unsuccessful connections in the (SIP,Dport) data set shown in Figure 3 are possible horizontal scans. Such streams can be detected automatically using good thresholding. However, IDGraphs allows an immediate deeper analysis. The suspect streams appear as several dark, splatted dots. By clicking on them, the user can reveal detailed textual information. In this case, we learn that all these streams are from the same SIP, and target different Dports: a vertical port scan. Since it is unlikely that the SYN-SYN/ACK failure count would be high for each of these streams if they each only addressed one DIP, the attack is likely also a horizontal scan, and therefore also probably a block scan.

Figure 4 highlights several suspicious streams. In particular, notice the two similarly shaped streams at the beginning of hour 0 (light green) and the beginning of hour 3 (light blue). Clicking on them, we find that they both communicate with Dport 3306, which is used by MySQL. These two possible attacks share the same temporal pattern; note especially the almost constant connection failure rate to the MySQL database for a time period of 15 to 20 minutes. We suspect this pattern may indicate a consistent hacking technique -- perhaps password guessing. By querying and selecting all the streams with this Dport, users can further examine all suspicious communication with MySQL in the dataset.(Figure 6)

### 5.3 SYN Flooding Pattern Discovery

Theoretically speaking, any streams with high SYN-SYN/ACK values in the (DIP,Dport) data set are potential TCP SYN flooding attacks. But IDGraphs allows users to pursue this initial hypothesis more deeply. Figure 11 reveals the temporal patterns of the most suspicious NetFlow streams, and shows that they had SYN-SYN/ACK values that peaked during hours 2 and 3. Brushing on a linked correlation matrix view (Figure 10) reveals four streams with very similar temporal patterns. Even though the DIPs and Dports for these streams are totally different, it is highly probable that these flooding attacks emanate from the same source. We tested this hypothesis by visualizing the same traffic keyed and aggregated by (SIP,DIP) (Figure 12). Querying for and highlighting streams with these four DIPs, we find that at any given time the the attacks indeed emanated from the same SIP. While SIPs did change over time, they were always from the same subnet. It seems the attacker was flooding destination hosts on a list, and trying to hide his attack by switching the SIP from time to time.

*Figures 11 and 12 about here*

### 5.4 Worm Propagation Pattern Discovery and Strategy Inference

Using IDGraphs time series based visualization, patterns in anomalous activity patterns are simple to spot. This offers clues about the propagation strategy of the associated attacks. For instance, we found a very regular series of periodic scans to TCP port 25 (servicing SMTP) as illustrated in Figure 13. It appears to result from the RTM Sendmail Worm[3]. The infected host sends out a burst of scan packets periodically, likely to avoid overloading the attacking machine and its network bandwidth.

*Figure 13 about here*

### 6 Comparison and Evaluation

In this section, we evaluate our IDGraphs system with comparisons to existing automated and visualization-based ID systems, and through a discussion with a practicing computer systems administrator.

---

[3] http://www.systemtoolbox.com/bfarticle.php?content_id=61, 2004.

### 6.1 Comparison with an automated IDS: HRAID

The High-speed Router-based Anomaly and Intrusion Detection (HRAID) system is a traditional IDS developed at Northwestern that uses automation to detect intrusions in real time [GLC05]. Its primary detection mechanism is the identification of network flows with unusually high failed connection counts. HRAID also features hash-table- ("sketch-") based data aggregation enabling low memory consumption and utility with high-bandwidth flows. IDGraphs was inspired by HRAID, and uses an intermediate HRAID netflow data format as input. IDGraphs also uses failure count as its primary detection measure, and like HRAID aggregates data with (SIP,Dport), (DIP, DPort), (SIP, DIP) keys. As a result, the same flows HRAID identifies as attacks are displayed at the top of the SYN-SYN/ACK axis in IDGraphs.

Nevertheless, there are several important differences between HRAID and IDGraphs. The first are the design goals. HRAID is targeting automated, real-time detection of network anomalies and attacks; rapid and accurate reaction to those attacks is the key to its success. We designed IDGraphs to complement HRAID by

- Providing *visual analysis and validation* of attacks and anomalies detected by HRAID. This should be particularly important as system administrators tune HRAID or other automated ID systems to their local networks, setting the statistical thresholds above which flows are treated as attacks and finding the optimal tradeoff between false negatives and positives.

- Capturing and *visualizing the temporal patterns* in anomalous network traffic. These temporal patterns are an underutilized component of attack signatures.

- *Exposing undetected attacks* that might be detected in HRAID using new heuristics, and providing the information needed to construct those heuristics.

Below we provide examples of each of these complementary functions.

*Visual analysis and validation*. Figure 14 shows the visual pattern of the top ten horizontal scans detected in HRAID in one day's NU netflow dataset. Mostly those are sudden bursts of (failed) connections, represented as isolated points in the IDGraph. Visualizations such as these might be used to tune automated detection thresholds.

*Figure 14 about here*

*Visualizing temporal patterns*. Statistical maxima are often not enough to identify attacks — attacks may spread their traffic across time and/or IPs, effectively "flying beneath the radar" of threshold-based detection schemes. For such attacks, alternative detection signatures must be found. Figures 13 shows attack that would be below many thresholds, and yet exhibit temporal patterns that may be useful in automated detection.

*Exposing undetected attacks*. In Figure 14, the dark block in hours 0 through 2 is well below threshold. Query reveals that it is formed by streams from multiple SIPs to ports 6129 and 1433 in common DIP subnets (Figure 15). It is highly possible those are backdoor types of attacking behaviors launched from infected or compromised hosts trying to exploit the vulnerability of other hosts. One might attempt to detect such an attack by summing the flows keyed by destination subnets and certain Dports over certain limited temporal and failure count ranges.

*Figure 15 about here*

### 6.2 Comparison to ID visualization systems

IDGraphs distinguishes itself from existing ID visualization systems in three primary respects. First, its use of a *failure count vs. time spatial mapping* makes traffic anomalies and their temporal signatures visually apparent. Second, its use of Histographs to *composite netflow streams* drastically improves the scalability of this approach. Finally, its use of a *wide range of netflow data fields* including SIP, DIP, Dport, time and SYN-SYN/ACK gives a very broad view of network activity.

#### *6.2.1 Comparison to PortVis*

PortVis [MMK04] visualizes netflow data using a Dport vs. Dport spatial mapping (two axes are mapped with complementary Dport subfields). Color at each location displays the amount of network traffic. Users can zoom in to improve reveal more information, including temporal netflow profiles and explicit Dport numbers and traffic levels. PortVis is useful for detecting anomalous activity at certain ports, but offers only limited support for identifying the most suspicious activity targeting such ports.

By aggregating data using (SIP,Dport) or (DIP,Dport) keys or querying by Dport, IDGraphs can also visualize traffic targeting reaching certain ports, though it cannot provide the same clear port-centric overview. Moreover, such views also make temporal and failure count information visible, can be linked to additional views keyed by (SIP,DIP), and clicked on to reveal stream-specific detail, facilitating follow up analysis.

#### *6.2.2 Comparison to VisFlowConnect*

VisFlowConnect [YYT04] employs parallel coordinates using three incoming, local and outgoing IP axes to show the netflows to and from a local subnet. Flow volume is mapped to line thickness, introducing some data occlusion problems, particularly with many flows or heavy network traffic

IDGraphs once more offers unique spatial mappings and access to a wider range of data, and also is much more scalable than VisFlowConnect. Use of failure count and time in the mapping allows users to segregate legitimate and normal network traffic quickly. These two data types and destination port information are always available to the user for detailed diagnosis. The data frequency-to-luminance mapping of Histographs offers a good solution to the occlusion problem faced by VisFlowConnect, making IDGraphs more scalable.

#### *6.2.3 Comparison to NVisionIP*

NVisionIP [LYL04] provides multiple views for visualizing netflow datasets. The main galaxy view uses a spatial mapping of IP addresses similar to the mapping for port used by PortVis: subnet on the horizontal axis, hosts on the vertical. Zooming to subnets and ultimately single machines is available for deep analysis.

IDGraphs doesn't provide as effective an overview of network activity by IP address as NVisionIP. However, users can easily highlight all the streams to or from a certain IP, and then segregate especially suspicious from more benign streams using the failure count axis. Temporal patterns can provide further indications of suspicious activity.

### 6.3 Practitioner Feedback

We demonstrated our system to a practicing network administrator, and received several comments and suggestions. To the administrator, the utility of IDGraphs in explaining the characteristics of identified attacks to managers and colleagues was obvious. He also confirmed that IDGraphs visualizations should be quite useful in validating and tuning automated ID tools.

In the future, the administrator would like to see IDGraphs working with live NetFlow data, perhaps by introducing a horizontal roll to the display. IDGraphs and automated ID tools such as HRAID might be more closely integrated, so that the likelihood that a flow is an intrusion trace might be displayed (instead of a simple binary threshold). Finally, the correlation function might be expanded, so that once a suspicious stream is identified, others "like" it might quickly be found.

### 6.4 Limitations

IDGraphs is certainly not without its limitations. Although we have reduced visual clutter our frequency-to-luminance mapping, clutter and occlusion are still quite evident when many streams are highlighted.(Figure 15) One solution we have already implemented uses a constant hue for highlighted streams and our regular frequency-to-luminance mapping to address occlusion. This does not, however, allow users to distinguish one highlighted stream from another.

We have already mentioned that although queries by individual SIP, DIP and Dport are available, our current spatial mapping does not provide a good overview of the data distribution in these individual fields. An improved overview might be provided by offering SIP, DIP and Dport slider that allow users "page" through the streams partially keyed by these data fields. Also, because Histographs can function with any spatial mapping, it might be interesting to experiment with SIP, DIP or Dport vs. time mappings.

Although rare, we have also found a few cases when many streams in a coordinated scan map to the same single pixel, but the because the pixel is not isolated in the display, splatting is ineffective, and the visualized streams do not stand out. It may be possible to make splatting take effect not only when the point is isolated not only in the spatially mapped dimensions, but in the luminance- (frequency-) mapped dimension as well

### 7 Discussion and Conclusion

IDGraphs is an interactive system for visualizing net flows, capable of detecting network anomalies and attacks including port scans, worm attacks, and SYN flooding. Perhaps more importantly, it can lead to useful insights concerning the propagation and intrusion patterns used in network attacks, even if they are distributed or spoofed.

While IDGraphs uses a time vs. SYN-SYN/ACK plot, most other ID visualization systems use plots based on IP address and/or port. Such address-based mappings are very useful, and the IDGraphs mapping between data and display space should complement them well. In future work, we plan to introduce views with different spatial mappings such as SIP vs. time. With linking and brushing, they should greatly increase the utility of IDGraphs. We are also working to make this system work with real-time, constantly changing data streams.

## 8 Acknowledgements

This research was supported by NSF grant 0093172. Our thanks to Peter Dinda for his suggestions, and to John Kristoff for his practitioner's viewpoint.

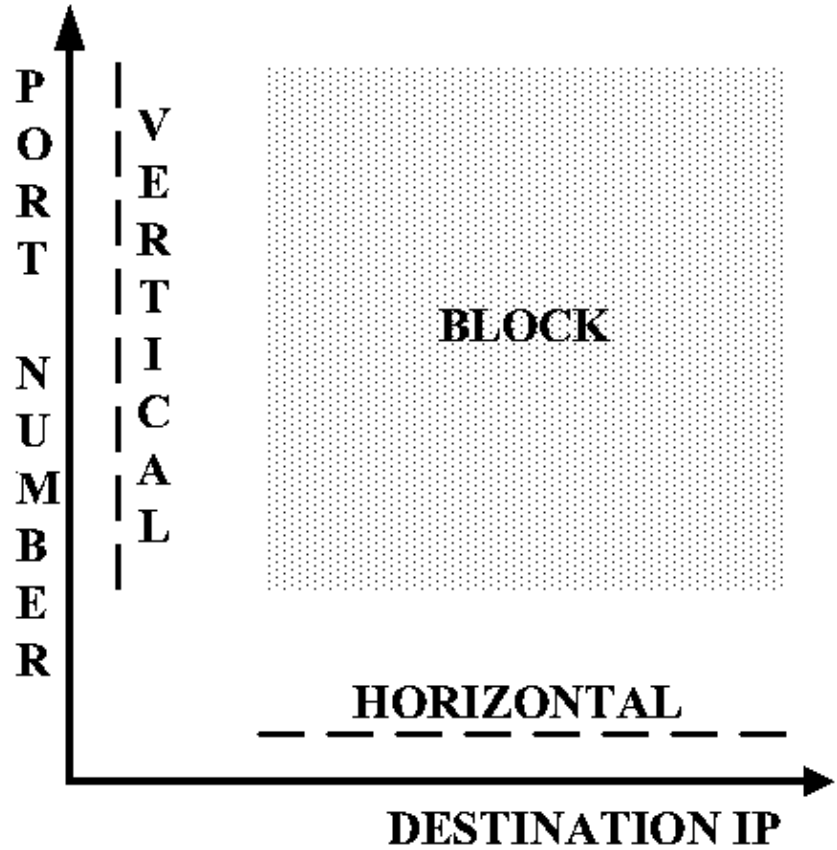

*Figure 1: Visual representation for three types of scans.*

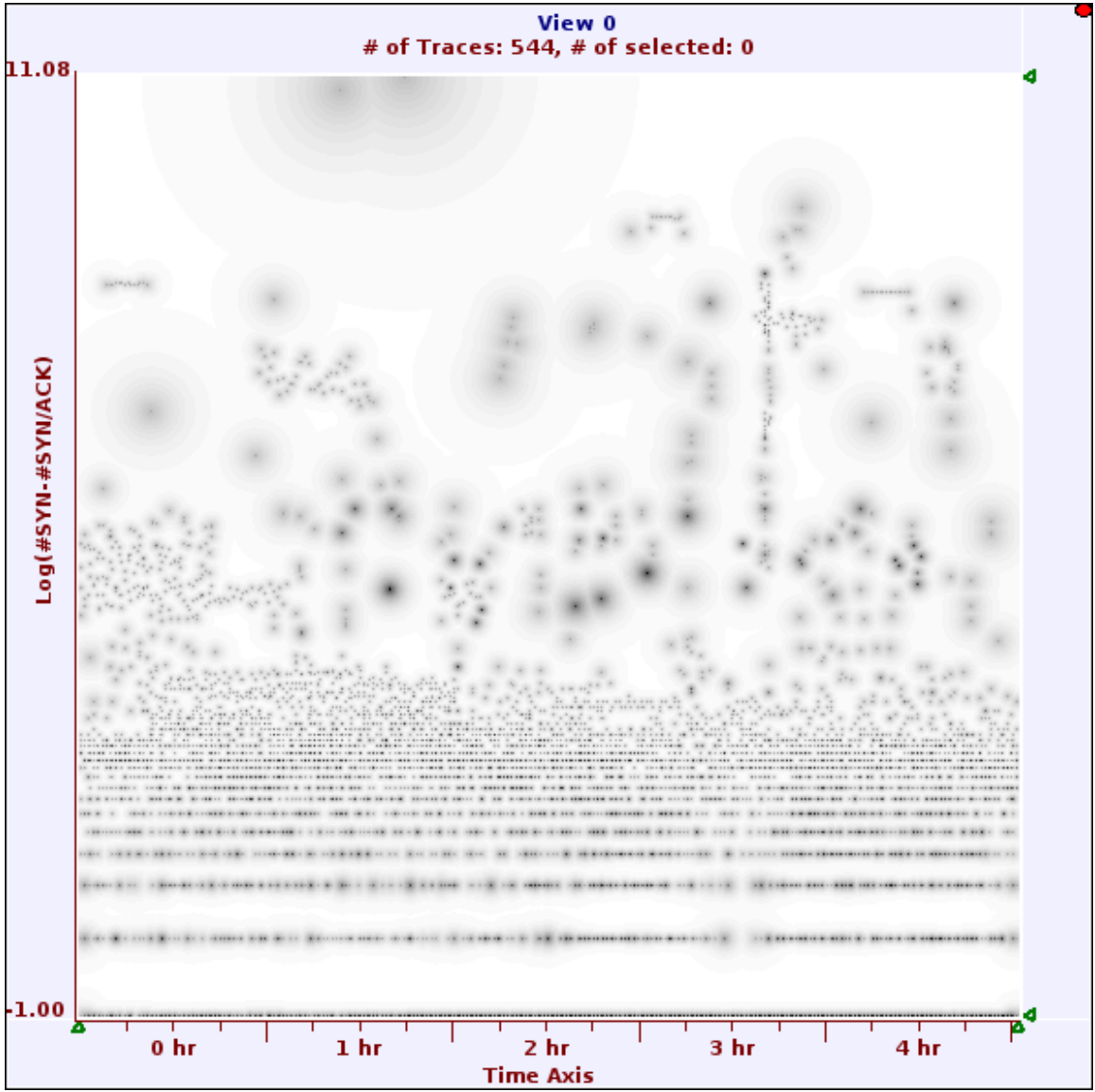

*Figure 2: (SIP,Dport) NetFlow streams plotted in a IDGraphs visualization todetect horizontal scans over many destination IPs. Dark points indicate high data density, and splatting (blurring) is used to increase the visibility of isolated points. Such unusually isolated and dark points attract attention, as do larger linear structures. Later query (Figure 3) and correlation analysis (Figure 9) reveals that the dark dots in hours 1 and 2 are block scans, while the linear structure in hour 3 is the outbreak of a worm attack with horizontal scans.*

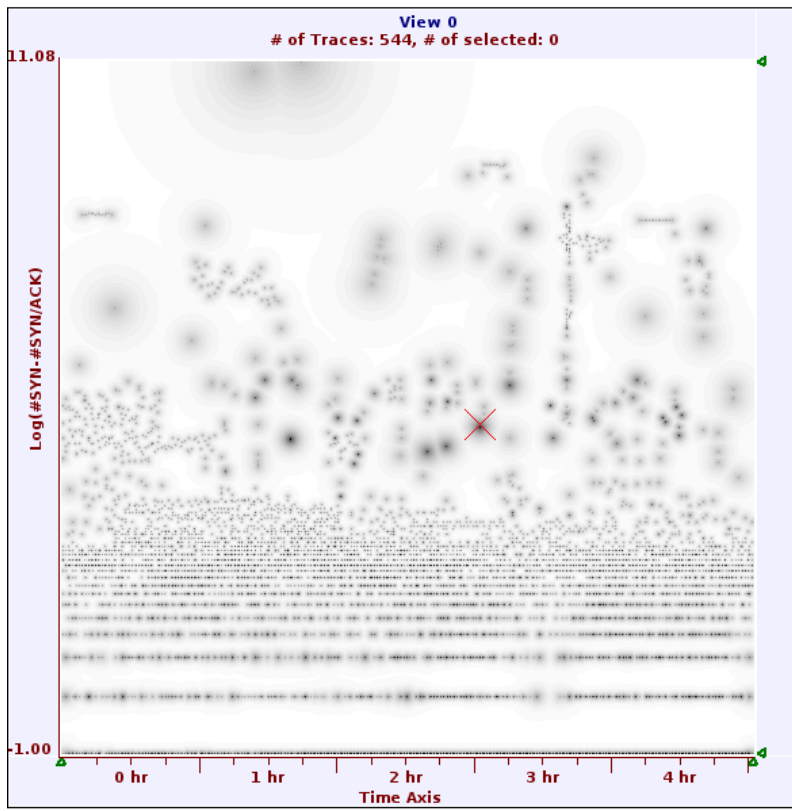

*Figure 3: The user clicks on one suspicious outlying and dark point (at the red X) in the (SIP,Dport) data to reveal the streams underneath it, in which a single IP scans multiple Dports -- a vertical scan.*

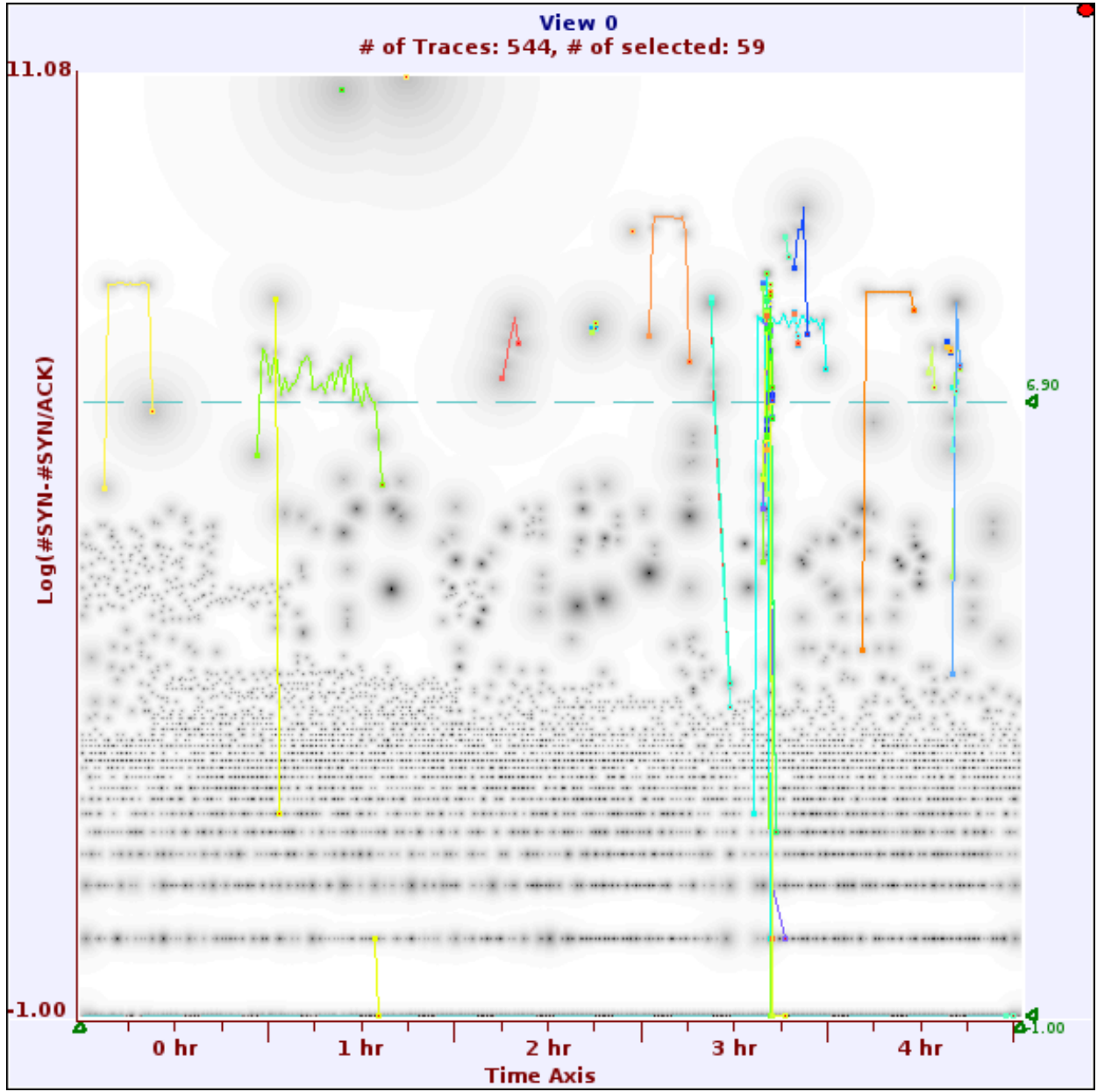

*Figure 4: Selection of streams in the (SIP,Dport) dataset with elements indicating more than 1000 unsuccessful connections (ln(1000)>=6.9). Plots points in the same graph are connected with line segments.*

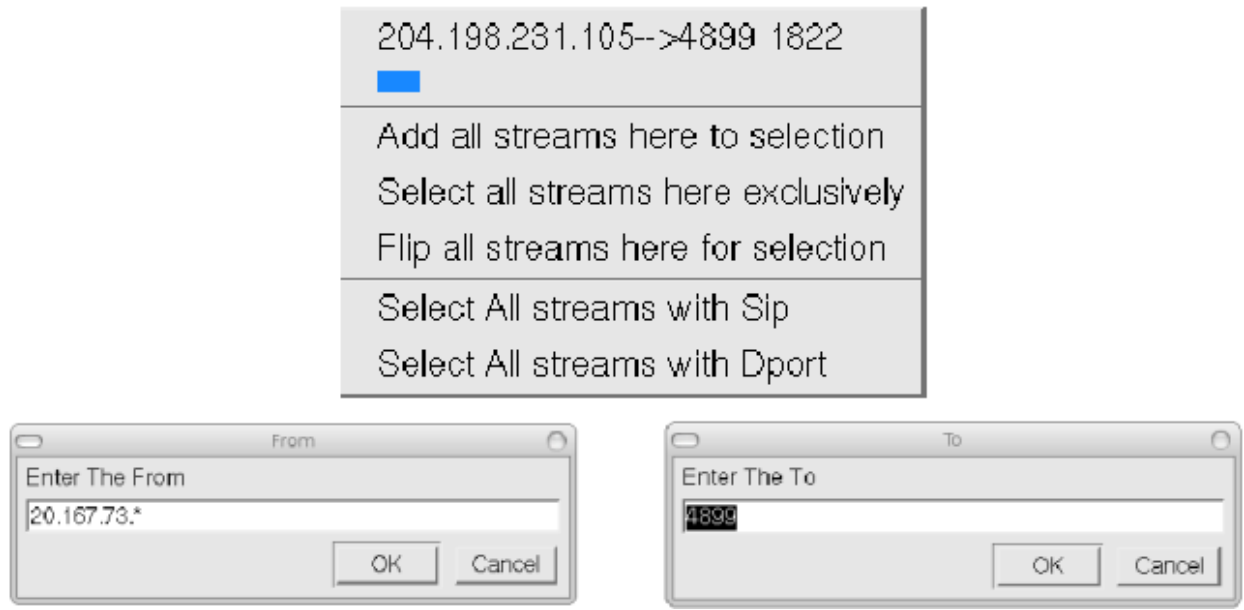

*Figure 5: The IDGraphs query interface allows users to select and highlight a subset of visualized streams by specifying SIP, DIP and/or Dport. Wildcards can be used to broaden the selection.*

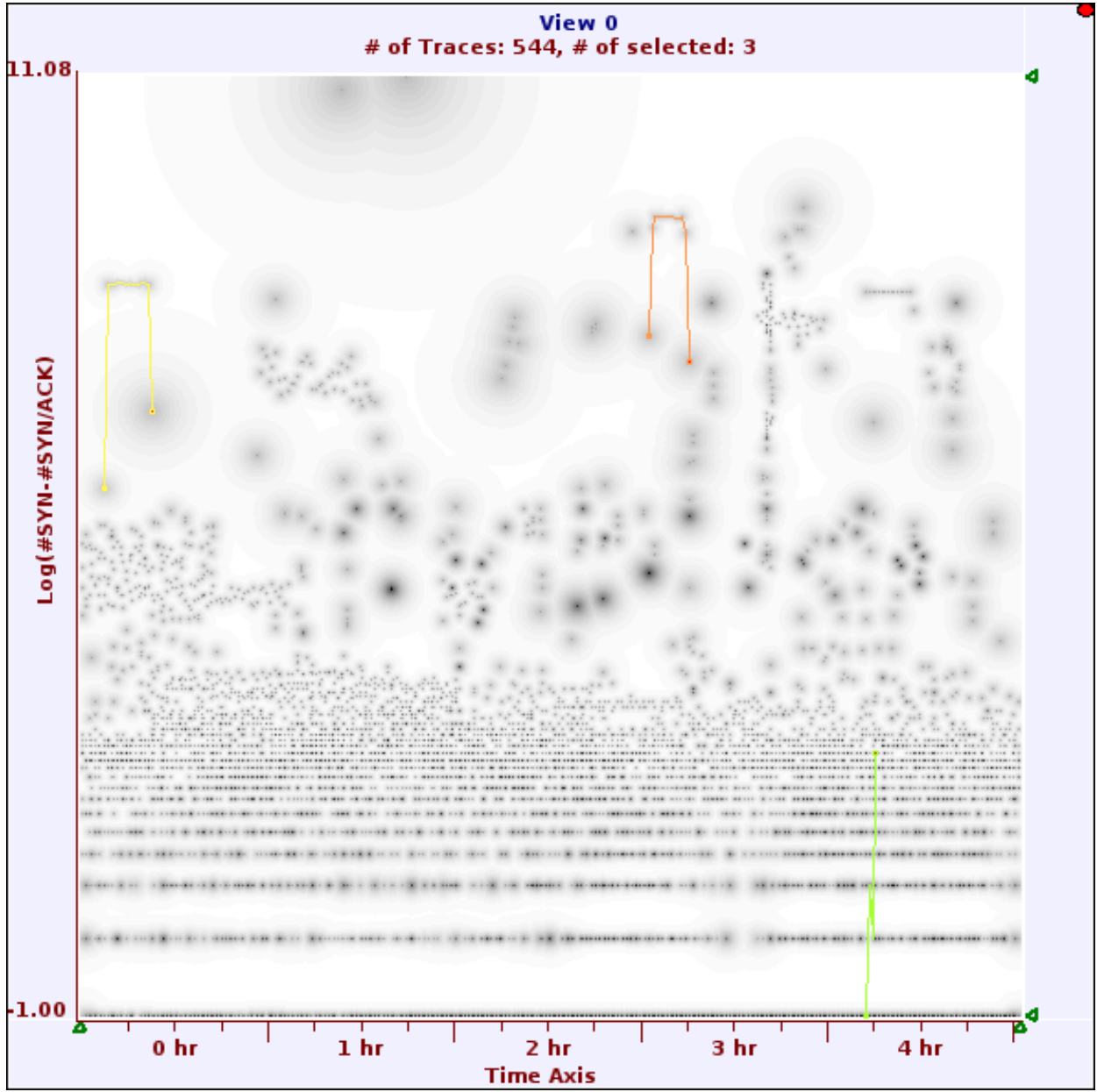

*Figure 6: Here the user has selected and highlighted all the streams with destination port 3306, which services MySQL.*

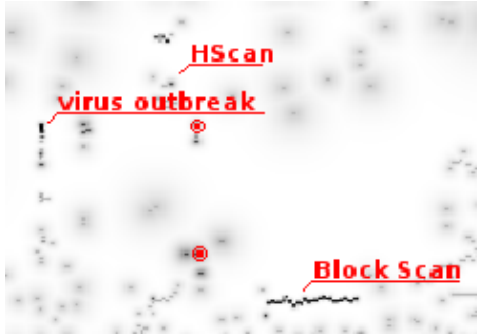

*Figure 7: IDGraphs allows user to annotate any point in the visualization. By default a red dot is left behind; clicking on it reveals the annotation text.*

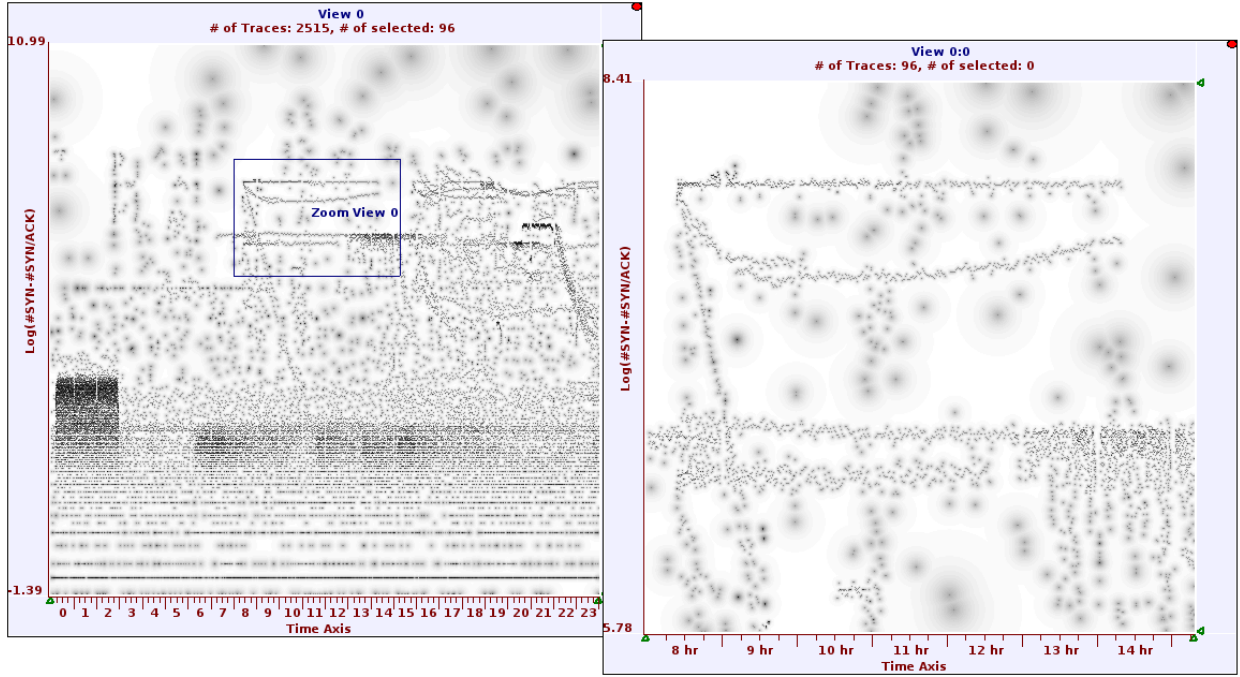

*Figure 8: Zoomed view to reveal data detail at a finer temporal scale.*

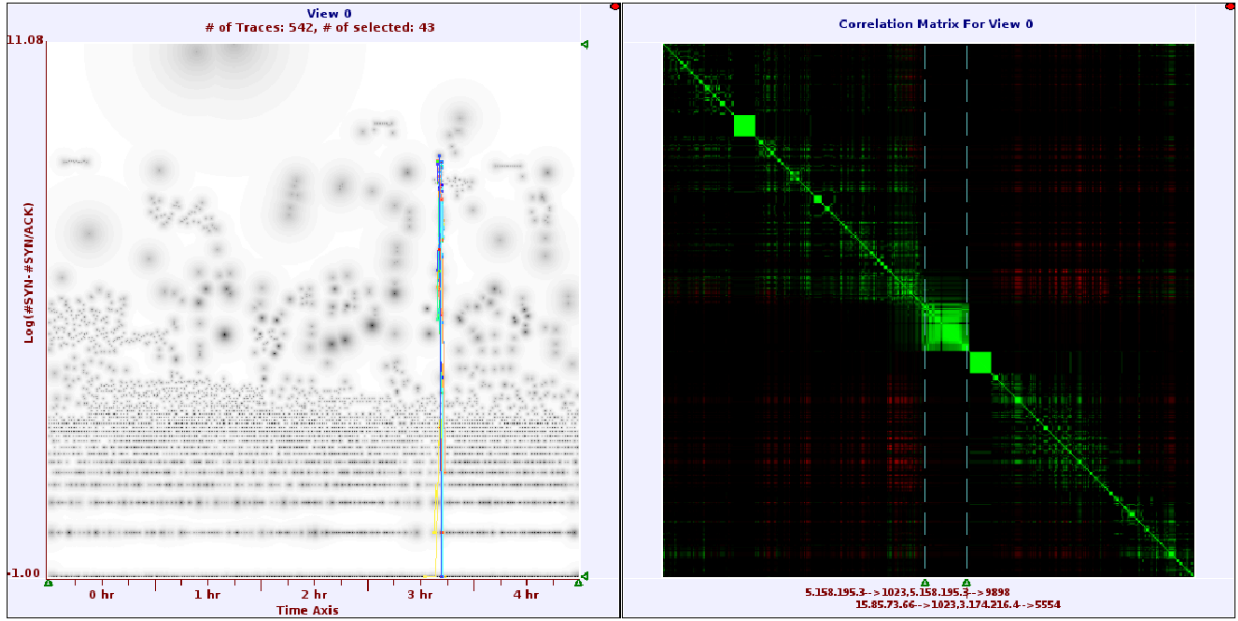

*Figure 9: Brushing with a linked correlation matrix view. Each row and column corresponds to one NetFlow stream. Green in a matrix cell indicates a positive correlation, red negative; brightness shows the magnitude of the correlation. We selected ("brushed") one highly correlated green block using the two horizontal sliders, the corresponding streams are then selected and highlighted in the main, linked IDGraph view. These highly correlated attacks are from different source hosts, targeting primarily three destination ports (a horizontal scan resulting from a worm virus attack).*

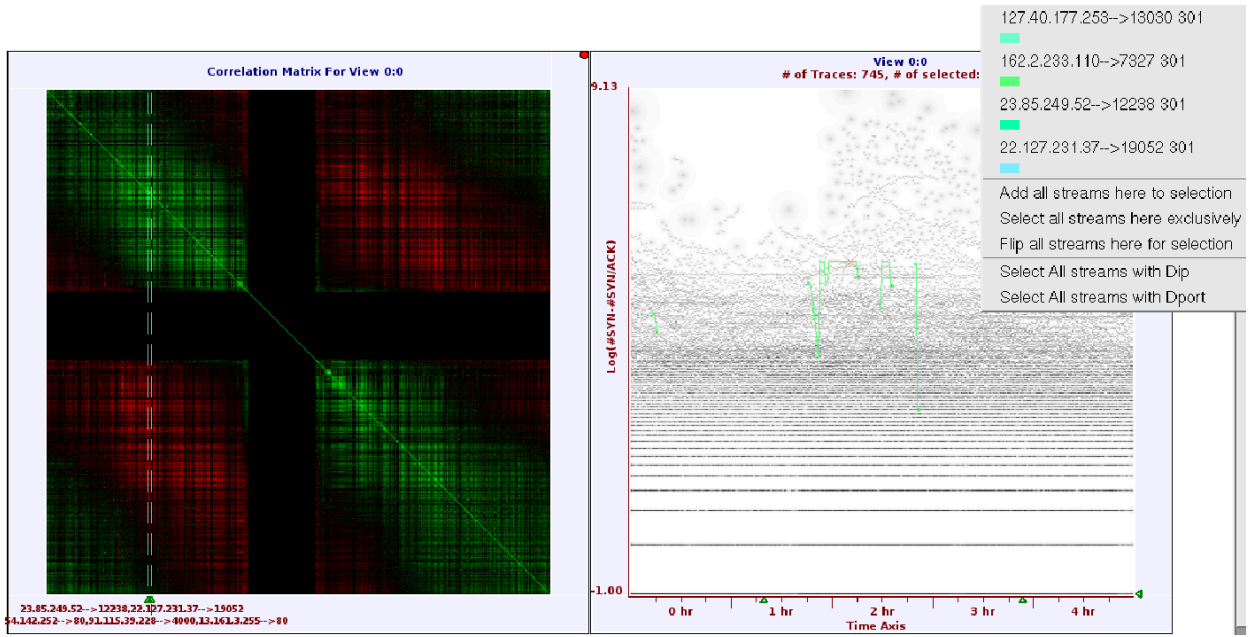

*Figure 10: Correlation brushing within a two hour time period. (Note the time slider in the main view to the right). Here we are visualizing a (DIP,Dport) input file to detect SYN flooding attacks. The four highly correlated streams selected here have the same pattern. Such parallel, coordinated attacks would be difficult to discover with traditional ID methods.*

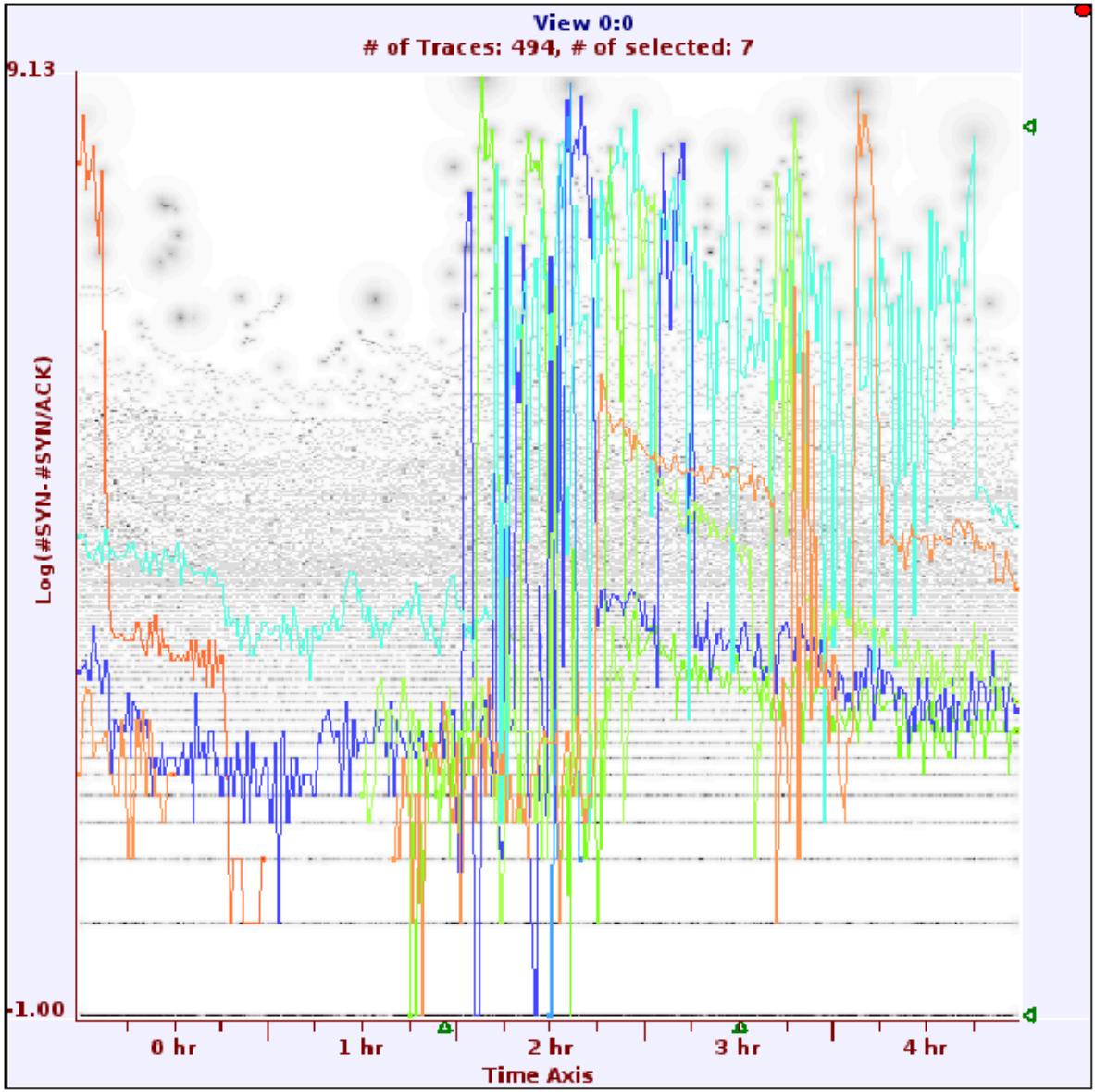

*Figure 11:The seven most suspicious sync flooding attacks selected and highlighted in a dataset key with (DIP,Dport)*

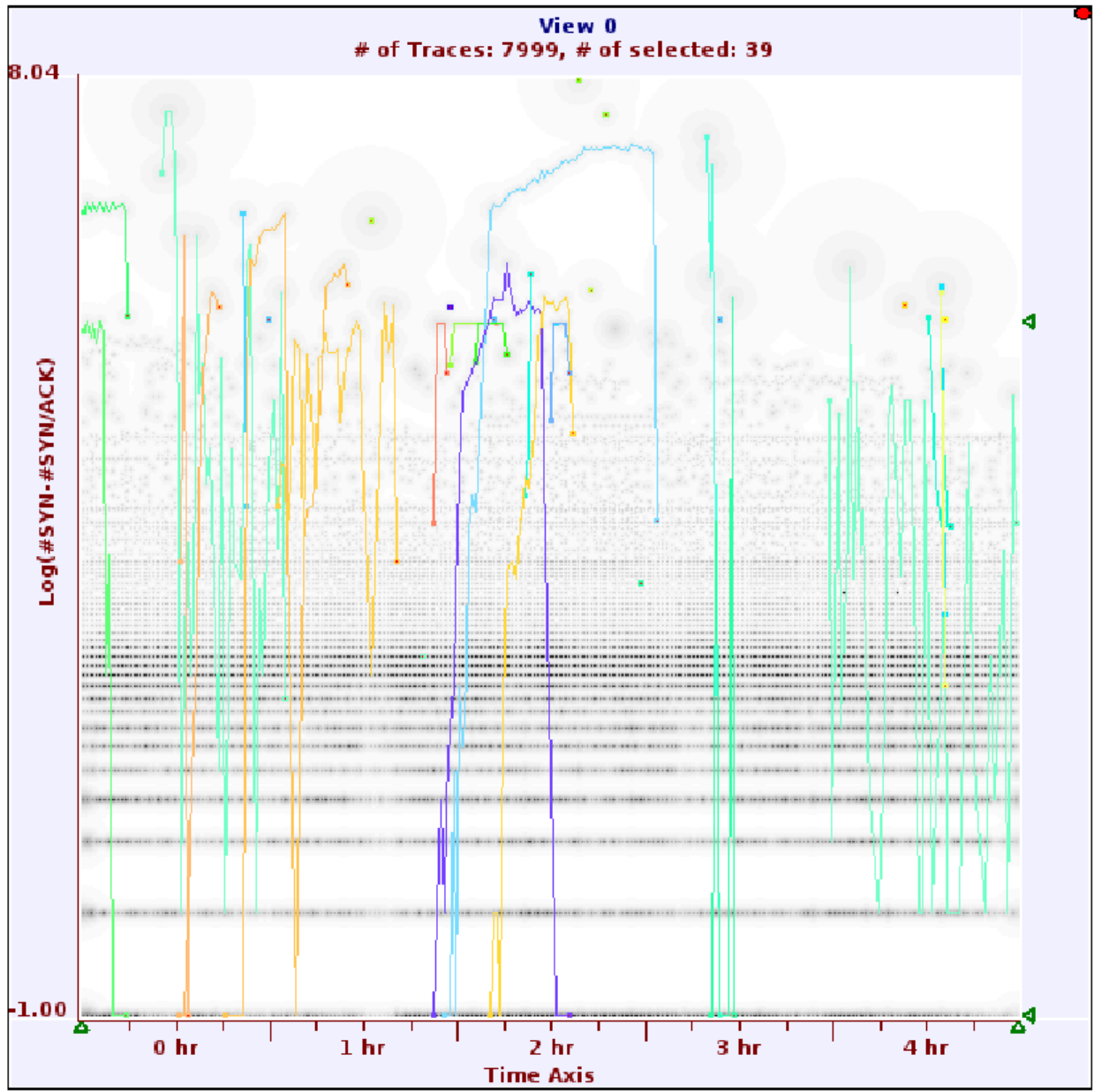

*Figure 12: The top suspicious potential attacks selected and highlighted in a dataset key with (SIP,DIP).*

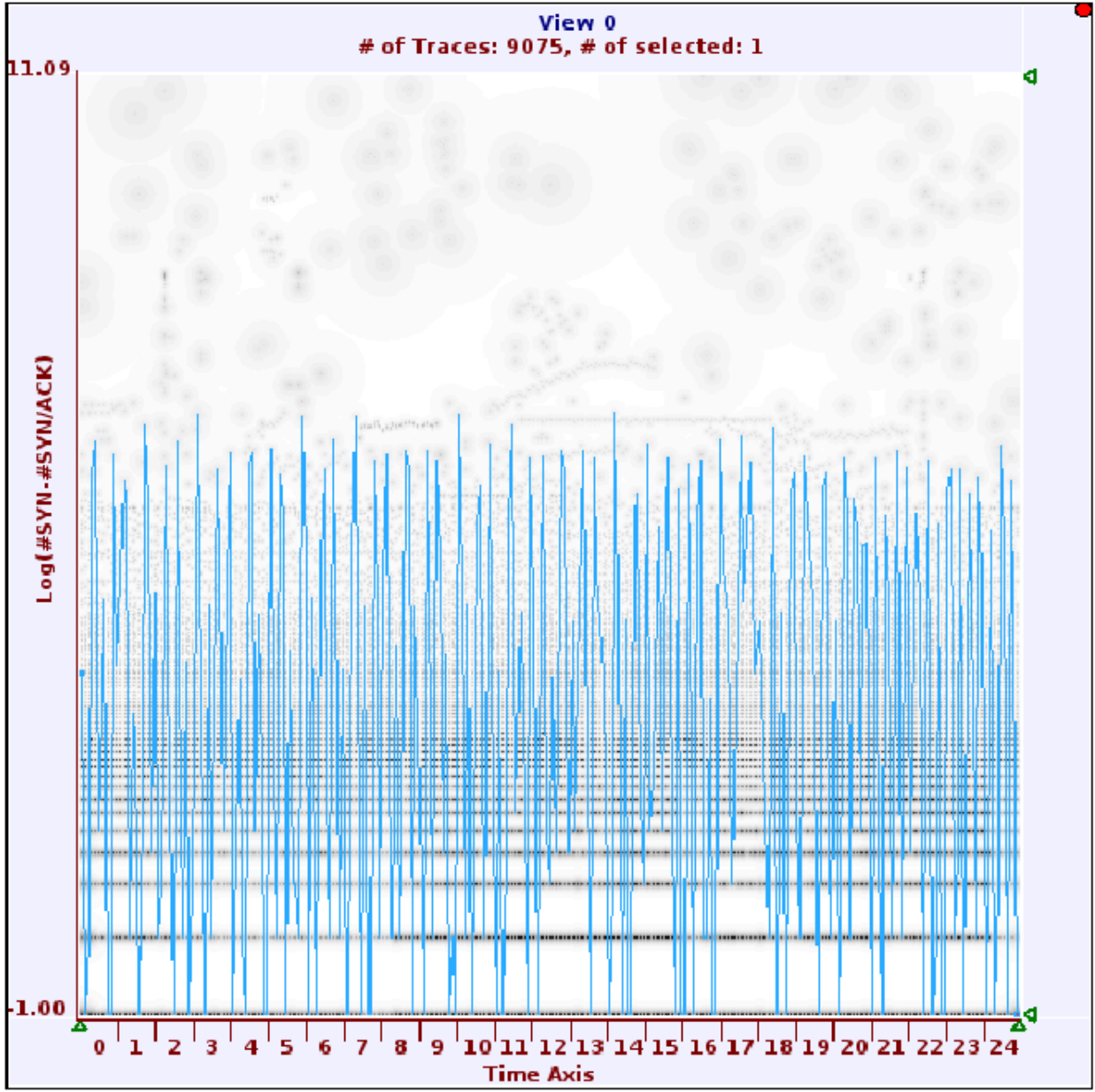

*Figure 13: Horizontal scanning revealed in the (SIP,Dport) data set. The highlighted stream shows a very obvious semi-periodic visual pattern over the graphed 25-hour period, with almost the same minimum (0) and maximum (~$800) SYN-SYN/ACK values for each burst.*

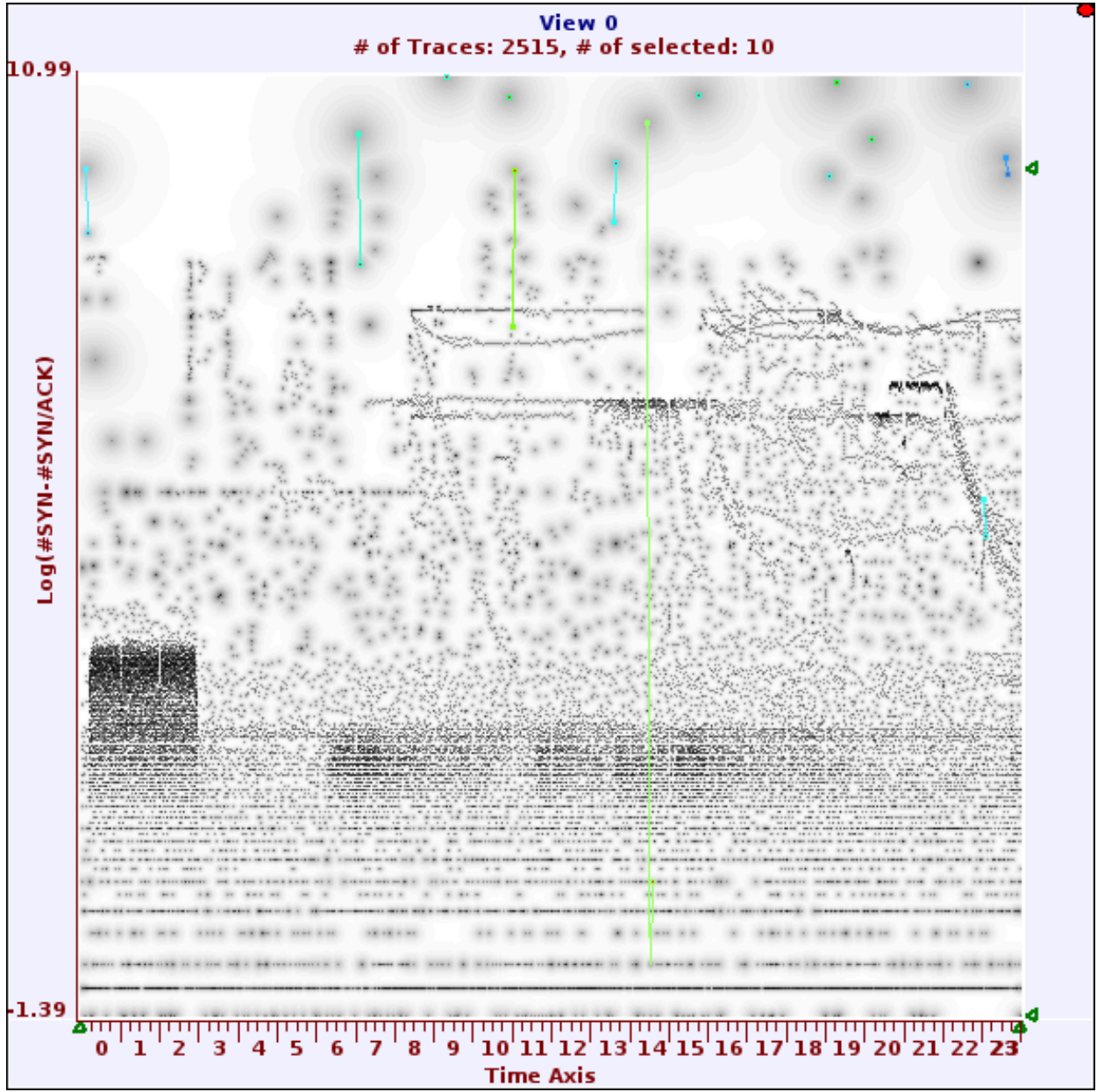

*Figure 14: Visual validation with detection results of HRAID. One day of data is aggregated by key (SIP,Dport), and then selected using a threshold slider. The 10 scans are the same detected and recorded in the HRAID log. Notice the dark rectangular region in the hours 0 to hour 2. Query results indicate that streams in that region are coordinated stealth attacks, and HRAID failed to detect them because the preset detection threshold is too high. We highlight those attacks in Figure 15.*

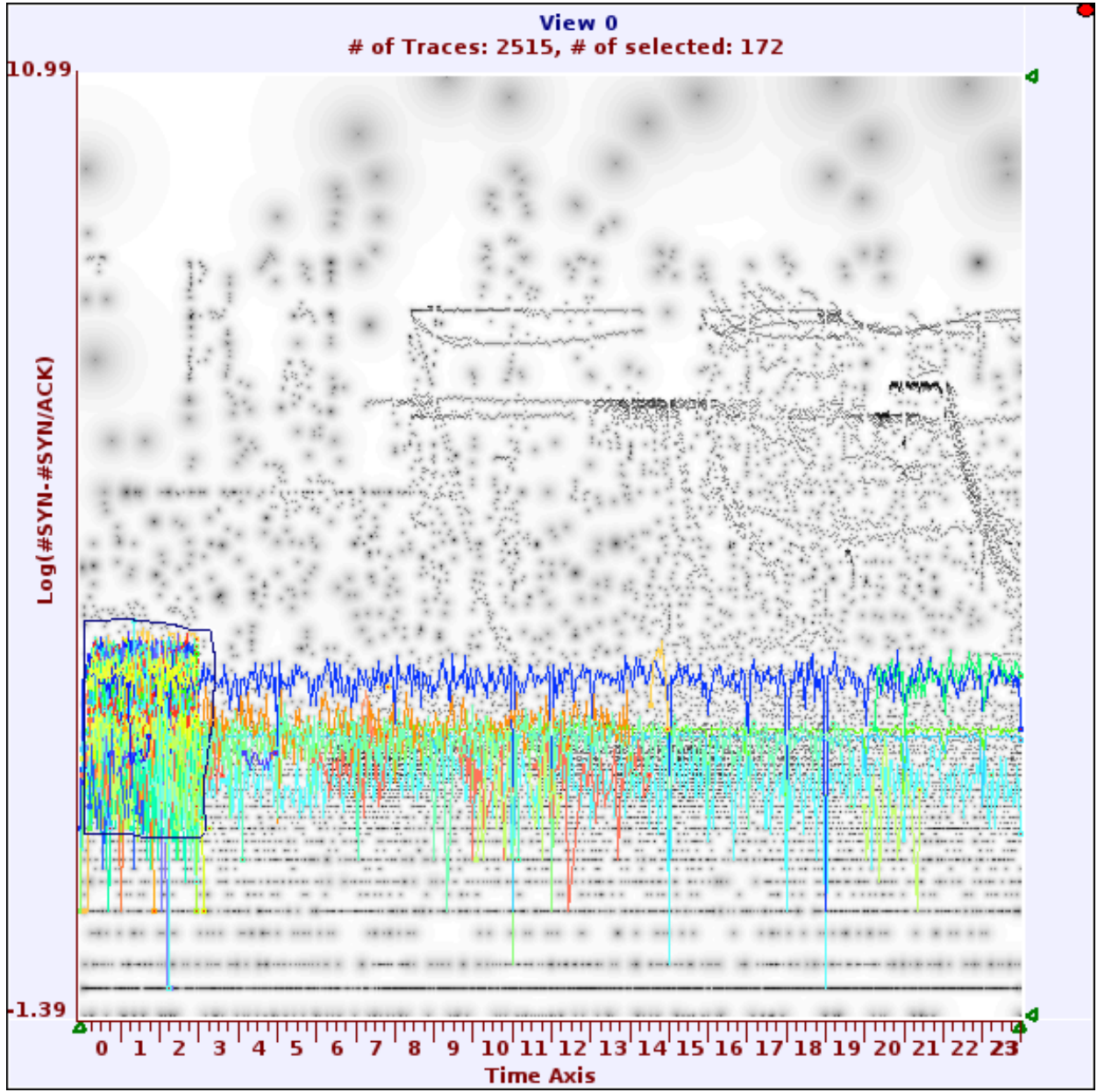

*Figure 15: Highlighting a coordinated stealth scan on port 6129 and 1433.*